\documentclass{sigchi-ext}
\usepackage[T1]{fontenc}
\usepackage{textcomp}
\usepackage[scaled=.92]{helvet} 
\usepackage{graphicx} 
\usepackage{balance}  
\usepackage{booktabs} 
\usepackage{ccicons}  
\usepackage{ragged2e} 

\usepackage[all]{nowidow}

\usepackage[utf8]{inputenc} 

\def\plaintitle{Dišimo: Anchoring our Breath}
\def\plainauthor{Jelena Mladenović, Jérémy Frey, Jessica R. Cauchard}
\def\plainkeywords{Biofeedback, Ambient display, Cardiac coherence}

\title{Dišimo: Anchoring Our Breath}

\numberofauthors{3}
\author{%
  \alignauthor{%
    \textbf{Jelena Mladenović}\footnote{Co-first authorship, both authors contributed equally to this work.}\\
    \affaddr{IDC Herzliya, Israel}\\
    \affaddr{Inria, France}\\
    \email{jelena.mladenovic@inria.fr} } \alignauthor{%
    \textbf{Jérémy Frey}\textsuperscript{\textit{a}}\\
    \affaddr{IDC Herzliya, Israel}\\
    \affaddr{Ullo, France}\\
    \email{jfrey@ullo.fr}} \\
    \vspace{0.25cm} \vfill \alignauthor{%
    \textbf{Jessica R. Cauchard}\\
    \affaddr{IDC Herzliya, Israel}\\
    \email{jessica.cauchard@idc.ac.il}} \\ \vspace{0.25cm}
 }

\definecolor{linkColor}{RGB}{6,125,233}
\hypersetup{%
  pdftitle={\plaintitle},
  pdfauthor={\plainauthor},
  pdfkeywords={\plainkeywords},
  bookmarksnumbered,
  pdfstartview={FitH},
  colorlinks,
  citecolor=black,
  filecolor=black,
  linkcolor=black,
  urlcolor=linkColor,
  breaklinks=true,
}

\copyrightinfo{Permission to make digital or hard copies of part or all of this work for personal or classroom use is granted without fee provided that copies are not made or distributed for profit or commercial advantage and that copies bear this notice and the full citation on the first page. Copyrights for third-party components of this work must be honored. For all other uses, contact the Owner/Author.\\
Copyright is held by the owner/author(s).\\
{\emph{CHI'18 Extended Abstracts}}, April 21-26, 2018, Montreal, QC, Canada.\\
ACM 978-1-4503-5621-3/18/04. \\
https://doi.org/10.1145/3170427.3186517}

\begin{document}

\teaser{
  \centering
  \includegraphics[width=1\textwidth]{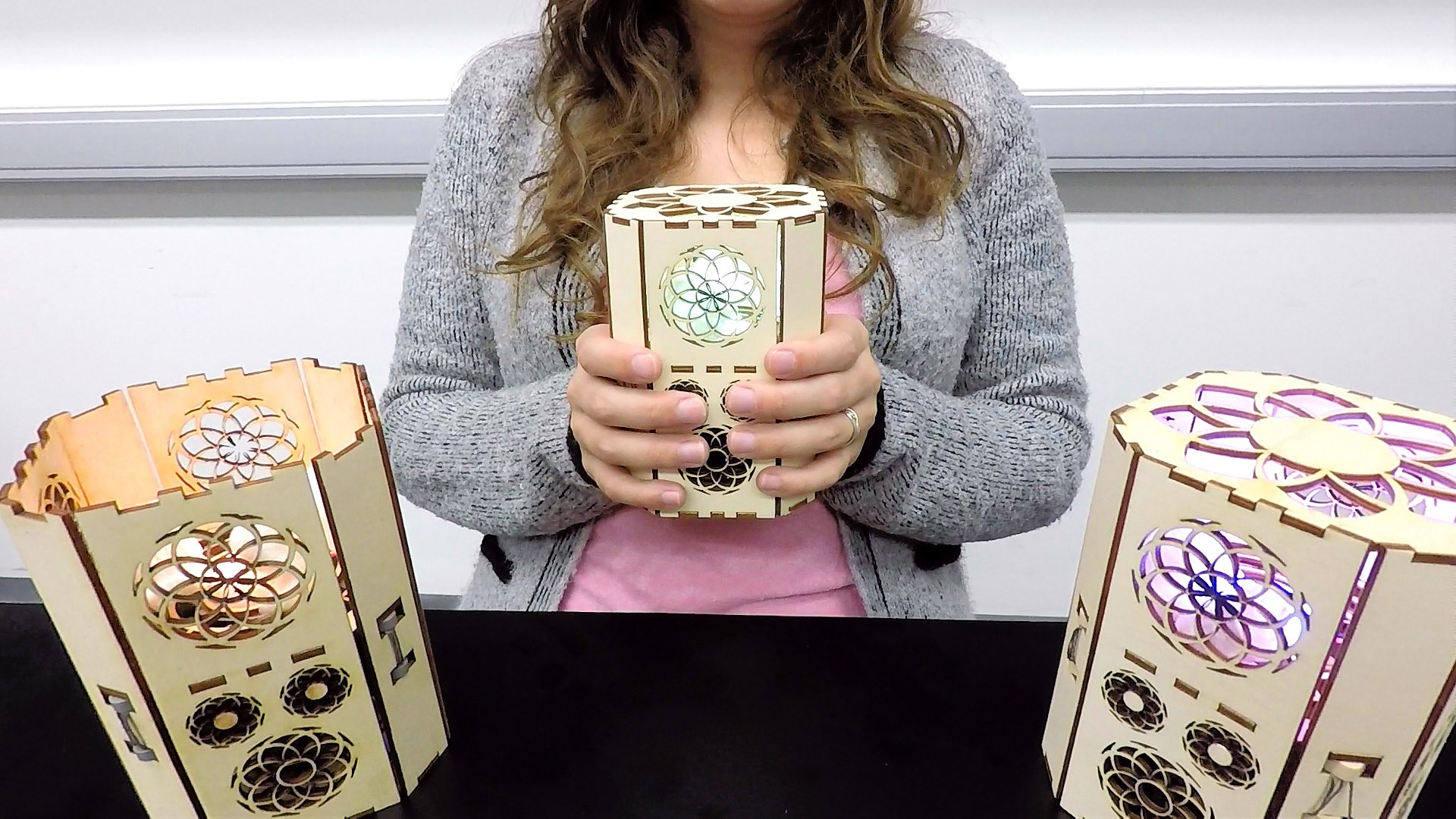}
  \caption{Dišimo: an ambient and shared biofeedback about heart rate variability.\label{fig:teaser}}
}

\maketitle

\RaggedRight{} 


\begin{abstract}

We present a system that raises awareness about users' inner state. Dišimo is a multimodal ambient display that provides feedback about one's stress level, which is assessed through heart rate monitoring. Upon detecting a low heart rate variability for a prolonged period of time, Dišimo plays an audio track, setting the pace of a regular and deep breathing. Users can then choose to take a moment to focus on their breath. By doing so, they will activate the Dišimo devices belonging to their close ones, who can then join for a shared relaxation session.

\end{abstract}

\keywords{\plainkeywords}

\category{H.5.2}{User Interfaces}{Interaction styles}
\category{H.1.2}{User/Machine Systems}{Human information processing}

\section{Introduction}

Over the last decades, the increasing availability of physiological sensors enabled new ways to mediate with the body. In human-computer interaction, while physiological activity has been used as another explicit input modality, researchers also investigated how presenting biofeedback to users could alter self-awareness and prompt for better habits. In \cite{Moraveji2011}, a widget on a desktop computer was used to help people breathe better. This feedback was effective, but was occurring somehow intrusively, and the locus of attention was still on a computer device. Recent works such as \cite{Roo2017} have started to explore how an ambient display with multimodal biofeedback could leverage physiological sensors to propose a calm experience focus on the body. In the present work, we take this idea further, incorporating real-time monitoring and exploring how a cluster of users can become an incentive for relaxation.

Dišimo ([dishimo], ``we breathe'' in Slavic languages) is a portable device that acts as a gentle and ambient reminder of one's state. It can be used not only to regulate oneself over the course of the day by breathing exercises, but several devices can be connected remotely in order to display those relaxation sessions to close ones.

We contribute to the field by leveraging behavior change with an ambient biofeedback which can be shared among a cluster. Not only the manifestation that a close one is using Dišimo could be an incentive to use the device and increase self-awareness, but joining a relaxation session could create an alternate way to empathize with other, by sharing explicitly physiological activity.

\section{Scenario}

When used in conjunction with a smartwatch capable of measuring heart rate, it will sense a decrease in heart rate variability (HRV) as a sign of stress or cognitive workload \cite{Fairclough2004}. After a certain period of time, if the HRV does not improve, Dišimo will play a sound as a gentle reminder to breathe. The user can then choose to take a break from the current task and use Dišimo to focus on breathing for his or her well-being. They can also choose to ignore it, in which case Dišimo will automatically stop playing the sound. If the person's state does not improve, after a little while, the sound will resume. In cases where users do not wear a heart rate sensor, the device could be connected to a computer and be triggered when the user stares at the monitor for too long.

If users decide to take a break and grasp Dišimo, the device illuminates itself and the sound will fade to let them breathe at their own pace. Then, if they manage to regulate their breathing and increase their HRV, physical particles embedded inside Dišimo will start to flutter and produce harmonious tones when hitting the enclosure. This later feedback serves as a physical manifestation of cardiac coherence, a state known to be correlated with well-being \cite{McCraty2009}. In future version, the device itself can be equipped with sensors to monitor heart rate by the mean of electrocardiography (ECG) when users grasp its edges.

During the demonstration, we propose to show how several devices could be used to orchestrate a co-located relaxation session. Up to 3 attendees will be able to use Dišimo. Their heart rate will be monitored, either by being equipped with a smartwatch or by direct contact with the device. Then, guided through audio with the modulation of a pink noise, within a couple minutes they will increase their HRV. Should an attendee manage to reach this state, their particles will flutter. Should all attendees increase HRV, the light of each device will brighten, an indicator of synchronization and shared relaxation.

\section{Description of the system}

\begin{marginfigure}[10pc]
  \begin{minipage}{\marginparwidth}
    \centering
  \includegraphics[width=1\columnwidth]{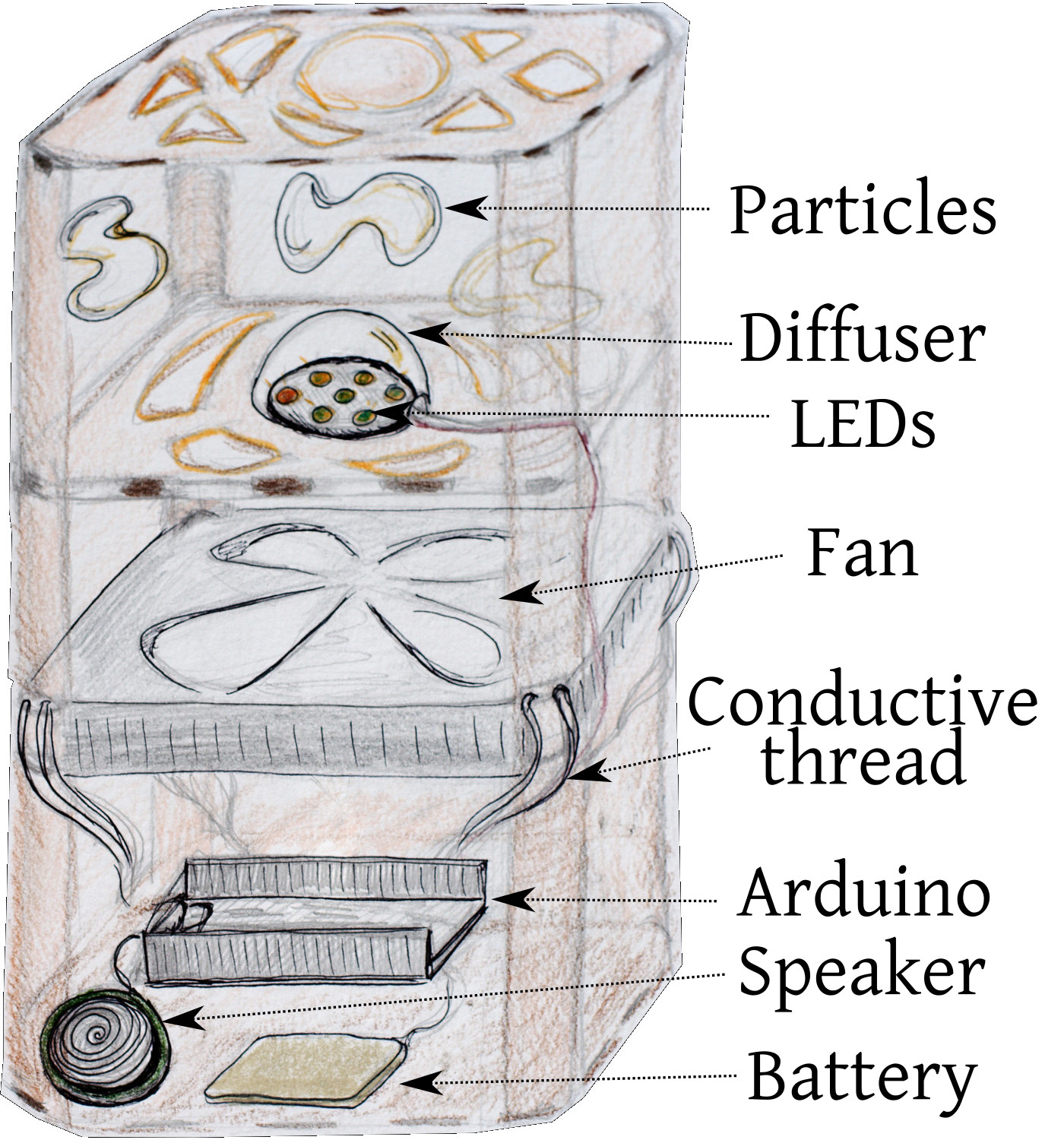}
  \caption{Dišimo Schematics: multimodal feedback is provided via sound, light, and fluttering particles.}~\label{fig:schema}
  \end{minipage}
\end{marginfigure}

Dišimo possesses two main features: it can sense the user and provide for a multi-modal feedback through light, sounds, and the actuation of physical particles. The bottom part of the device comprises its electronic components: an Adafruit Feather with a custom shield to power a speaker and detect when a user is grasping the device through capacitive touch -- a conductive thread being woven around the edges (Figure \ref{fig:schema}). A step-up voltage regulator is also present in order to power up a 12V fan (Noctua NF-A8/R8 PWM) which is enclosed in the center of the device.

\begin{figure}[!h]
  \centering
  \includegraphics[width=0.9\columnwidth]{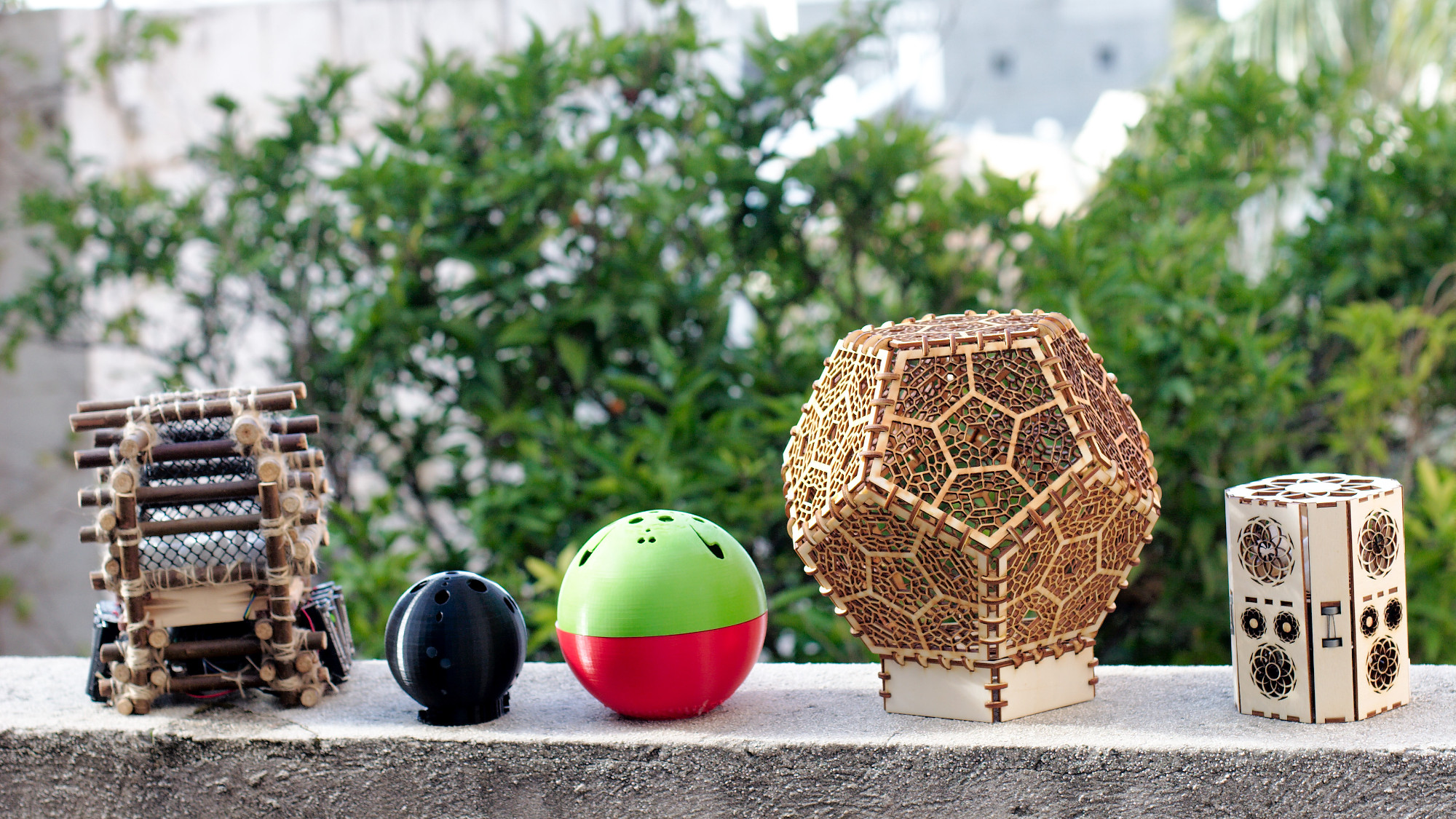}
  \caption{Various iterations were necessary in order to find a material which would be pleasing to hold (wood vs plastic), a structure that lets enough air flow toward the particles, and a shape which would create harmonious sound when the latter are fluttering (faces vs curves).}~\label{fig:iterations}
\end{figure}

The upper part of the device embeds LEDs (Adafruit Jewel) and a dome to diffuse their light. The upper section also serves as a ``chamber'' where particles can bounce when the fan is activated. Those particles are made of expanded polystyrene (Storopack Pelaspan). While they were meant to fill boxes and protect goods, they are light enough to fly with a modest airflow (from $\approx 34.8 m^{3}/h$) and, more importantly, because of their material and shape, they happen to create a nice sound when they hit a surface, a sound reminiscent of wooden wind chimes. To provide for another sort of biofeedback, that is the manifestation of air flowing, is one of the core design idea of the project. The device underwent several iterations in order to find the right combination of shape and material in order to accommodate this particular type of feedback, which is triggered once HRV increases (Figure \ref{fig:iterations}). 

The Adafruit Feather board can connect to a host computer through Bluetooth BLE in order to retrieve HRV when the user is wearing a device capable of measuring heart rate, such as the Mio Alpha 2 (used in the current prototype). It can also be paired with an OpenBCI board, an amplifier dedicated to the recording of physiological signals such as ECG. The OpenBCI board fits inside the device and is sensitive enough to measure ECG upon contact with the conductive material placed around the edges of Dišimo, which then serve as electrodes connected through a bipolar montage.

Whenever measurements are made continuously with a smartwatch or discreetly with the OpenBCI board, signals are processed in real-time on a host computer running OpenViBE. To asses HRV, we measure the range of instantaneous heart rate over a 15s sliding time window, with thresholds $HRV_{low}$ < 2 beats/min and $HRV_{high}$ > 5 beats/min. Upon detection of a low HRV for 10m, the breathing guide played through the speaker is a modulated pink noise, as in \cite{Roo2017}, which reminds of the sound of waves. The synthesized breathing is ample and slow enough to induce an increase in HRV, with a frequency of 7.5 breaths per minute. We also used the results from \cite{Frey2018} and increased the amount of time spent exhaled since such breathing feature was associated with positive emotions. The resulting sequence is $3\frac{1}{3}s$ breathe in, $3\frac{1}{3}s$ breathe out and finally a pause of $1\frac{1}{3}s$, that is repeated to form the guiding pattern.

The audio guide is played for 30s (4 breaths), after which the device will go silent. The volume of the audio guide is low enough so as not to compete with the attention level of users, avoiding being a push notification which would disturb them. It is eventually up to the user to decide whether they want to take a break or not. If they decide to use the device, upon grabbing, the audio guide will play again for 30s. It is not played continuously as the synthesized pattern only serves as example, and each user might have a slightly different pace of breathing.

When the device is touched, it is formally activated and it lights up with a color previously picked by the user. Among a cluster of users, the color of active users are mixed together (average of each RGB channel). The global brightness of the light is mapped to the ratio of active users who increased their HRV. In order to be as non-judgmental as possible, we purposely avoided to give information about who specifically reached a higher HRV. Dišimo is not meant to foster competition, that would defeat the purpose of improving well-being, instead it is an aid and a mediator.

\section{Audience and Relevance}

The demonstration is suitable for any audience: students, researchers, naive or expert in physiological computing. Most existing feedback about one's physiology are either personal, based on explicit metrics or compete with users' attention. Through this demonstration, we would like to advocate for the use of ambient biofeedback that can be shared among close ones. By trying the system first-hand, attendees will be able to reflect on the topic and experience a group relaxation session. We expect Dišimo will foster discussions about issues related to physiological computing, such as privacy or how to determine which (inner) state should be desired, if any.

We would gladly share the tools and methodology used to craft Dišimo with peers interested in leveraging well-being through computing. We believe that such interfaces, which react to users' states and are anchored in reality, could promote self-awareness.	

\balance{}

\bibliographystyle{SIGCHI-Reference-Format}
\bibliography{biblio}

\end{document}